Michał WANIC[1]

# MAGNETOELECTRIC MULTIFERROICS: FROM FUNDAMENTALS TO TRANSFORMATIVE APPLICATIONS – A MINI REVIEW

**Abstract:** Multiferroics, combining ferroelectric and magnetic orders, enable magnetoelectric (ME) coupling for advanced applications. This mini review explores single-phase and composite multiferroics, examining phenomenological, microscopic, nanostructured, and quantum mechanisms driving ME effects. Phenomenological models quantify coupling coefficients, while microscopic approaches reveal spin-lattice interactions, including frustrated spin states and Dzyaloshinskii-Moriya contributions. Nanostructured systems, such as plasmonic skyrmion lattices and metasurfaces, enhance ME effects for tunable birefringence and electromagnon amplification. Quantum heat engines utilize spin entanglement and topological protection in chiral chains and skyrmion lattices for efficient energy conversion. Applications include high-sensitivity magnetic sensors, tunable radio-frequency devices, energy-efficient MERAM, energy harvesters, quantum heat engines, and thermal diodes. Future research aims to optimize room-temperature ME coupling, scalability, coherence, and biocompatibility for innovations in sensing, quantum computing, and sustainable energy.

## 1. Introduction

The magnetoelectric effect was theoretically predicted by I. E. Dzyaloshinskii in 1959 through an analysis of the crystal lattice symmetry in $Cr_2O_3$ [1]. These theoretical considerations were experimentally validated in the early 1960s by D. N. Astrov, as well as by V. J. Folen, G. T. Rado, and E. W. Stalder [2, 3]. These studies marked the beginning of research interest in single-phase multiferroics during the 1960s and 1970s. However, research progress slowed due to the low magnetoelectric coupling coefficients, which were typically observed only at very low temperatures, and due to technological limitations in synthesizing single-phase materials exhibiting both magnetic and electric ordering. The lack of satisfactory properties at room temperature significantly restricted the practical applications of multiferroics, leading to a decline in scientific interest.

---

[1] Corresponding author: Michał Wanic, Department of Physics and Medical Engineering, Faculty of Mathematics and Applied Physics, Rzeszow University of Technology, Powstańców Warszawy 6, 35-959 Rzeszow, Poland, e-mail: mwanic@prz.edu.pl

Multiferroics, materials that simultaneously exhibit ferroelectricity and ferromagnetism (or other magnetic ordering), have experienced a renewed surge in interest since the early 21st century, driven by several factors [4, 5, 6, 7]. Advances in the synthesis of single-phase monocrystals have led to the discovery of new types of multiferroics with novel ferroelectricity mechanisms, such as those driven by spin-driven polarization [8]. A defining feature of multiferroics is the magnetoelectric (ME) coupling, which enables the control of magnetization by an electric field and, conversely, the manipulation of electric polarization by a magnetic field, opening pathways for energy-efficient devices. Concurrently, the development of numerical methods, such as density functional theory, has facilitated the design of new materials and the analysis of factors influencing magnetoelectric coupling. Additionally, progress in thin-film growth techniques has enabled the practical implementation of numerical analyses, enhancing control over material properties at the nanoscale. This resurgence of interest is driven not only by technological advancements in multiferroic synthesis but also, and perhaps primarily, by their potential applications. These include magnetoelectric-based random-access memory (RAM) devices, magnetic field sensors, tunable radio-frequency devices, energy harvesting systems, thermal energy transfer, and a wide range of applications in biomedical technologies. This mini review first explores magnetoelectric phenomena in multiferroics through phenomenological, microscopic, nanostructured, and quantum approaches, followed by a discussion of their classification into single-phase and composite materials and applications in memory devices, sensors, quantum heat engines, and energy technologies.

## 2. Magnetoelectric Phenomena in Multiferroics

Multiferroic materials, exhibiting simultaneous ferroelectric and magnetic orders, enable unique magnetoelectric coupling for advanced applications. This chapter explores phenomenological, microscopic, nanostructured, and quantum approaches to understanding these phenomena. Each perspective highlights distinct mechanisms driving energy transfer and device functionalities.

### 2.1. Phenomenological Approach

Multiferroic materials exhibit simultaneous ferroelectric and magnetic ordering, enabling unique magnetoelectric coupling, where electric fields influence magnetization and magnetic fields induce electric polarization. The phenomenological description of these phenomena relies on Landau's theory, which models the free energy of a homogeneous magnetoelectric system as a function of macroscopic order parameters: spontaneous polarization $(\vec{P_S})$ and magnetization $(\vec{M_S})$. This approach provides a framework to quantify ME coupling without delving into microscopic mechanisms, focusing on thermodynamic behavior and symmetry constraints [4, 7].

The free energy $F(\vec{E}, \vec{H})$ of a multiferroic system at constant temperature is approximated by a Taylor expansion in terms of the electric field $(\vec{E})$ and magnetic field $(\vec{H})$, as described by:

$$F(\vec{E}, \vec{H}) = F_0 - P_{S,i} E_i - M_{S,i} H_i - \frac{1}{2}\epsilon_0 \epsilon_{ij} E_i E_j - \frac{1}{2}\mu_0 \mu_{ij} H_i H_j - \alpha_{ij} E_i H_j - \frac{1}{2}\beta_{ijk} E_i H_j H_k - \frac{1}{2}\gamma_{ijk} H_i E_j E_k - \cdots \quad (1)$$

Here, $F_0$ is the ground state free energy, $P_{S,i}$ and $M_{S,i}$ are components of spontaneous polarization and magnetization, respectively, $\epsilon_0$ and $\mu_0$ are the dielectric and magnetic susceptibilities of vacuum, $\epsilon_{ij}$ and $\mu_{ij}$ are second-order tensors of dielectric and magnetic susceptibilities, and $\alpha_{ij}$, $\beta_{ijk}$, and $\gamma_{ijk}$ are coupling coefficients. The subscripts $(i, j, k)$ denote spatial coordinates. The term $\alpha_{ij} E_i H_j$ represents the linear ME coupling, while higher-order terms $(\beta_{ijk}, \gamma_{ijk})$ describe nonlinear contributions.

The ME effect quantifies the coupling between electric and magnetic fields in matter. By differentiating the free energy with respect to the electric field, the total electric polarization is obtained:

$$P_i(\vec{E}, \vec{H}) = -\frac{\partial F}{\partial E_i} = P_{S,i} + \epsilon_0 \epsilon_{ij} E_j + \alpha_{ij} H_j + \frac{1}{2}\beta_{ijk} H_j H_k + \gamma_{ijk} H_i E_j + \cdots \quad (2)$$

In the absence of an electric field $(\vec{E} = 0)$, the magnetic field-induced polarization (direct ME effect) is:

$$P_i = \alpha_{ij} H_j + \frac{1}{2}\beta_{ijk} H_j H_k \quad (3)$$

Similarly, differentiating with respect to the magnetic field yields the total magnetization:

$$M_i(\vec{E}, \vec{H}) = -\frac{\partial F}{\partial H_i} = M_{S,i} + \mu_0 \mu_{ij} H_j + \alpha_{ij} E_j + \beta_{ijk} E_i H_j + \frac{1}{2}\gamma_{ijk} E_j E_k + \cdots \quad (4)$$

In the absence of a magnetic field $(\vec{H} = 0)$, the electric field-induced magnetization (converse ME effect) is:

$$M_{S,i} = \alpha_{ij}E_j + \frac{1}{2}\gamma_{ijk}E_jE_k \tag{5}$$

The tensor $\alpha_{ij}$ governs the linear ME effect, describing the induction of polarization by a magnetic field or magnetization by an electric field. Higher-order tensors $(\beta_{ijk}, \gamma_{ijk})$ capture nonlinear ME interactions, which become significant in strong fields or specific material symmetries.

### 2.2. Microscopic Approach

While phenomenological models describe magnetoelectric (ME) coupling using macroscopic variables, microscopic approaches reveal the atomic and electronic mechanisms driving this behavior. Group theory provides symmetry-based criteria for ME effects but cannot predict their magnitude. Understanding the microscopic origins of coupling coefficients, such as $\alpha_{ij}$, requires analyzing interactions between spins, ions, and fields at the lattice level [9, 10, 11]. Key microscopic mechanisms include:

- Single-Ion Anisotropy ($\propto (S_i^\alpha)^2$): An applied electric field displaces ions relative to their ligands, altering the ligand field's anisotropy, magnitude, or symmetry, which influences local magnetic properties.
- Symmetric Superexchange $\left(\propto r_{ij}\left(S_i^\alpha S_j^\beta + S_i^\beta S_j^\alpha\right)\right)$: Electric fields modify ion positions and electron wave functions, changing orbital overlaps and exchange integrals, thus affecting magnetic ordering.
- Antisymmetric Superexchange $\left(\propto r_{ij}\left(S_i^\alpha S_j^\beta - S_i^\beta S_j^\alpha\right)\right)$: Dzyaloshinskii-Moriya (DM) interaction induces spin canting, contributing to ME coupling. Though weaker than symmetric exchange, its modification by electric fields can be significant.
- Dipolar Interactions $\left(\propto \frac{\vec{m_i}\cdot\vec{m_j}}{r_{ij}^3} - \frac{3(\vec{m_i}\cdot\vec{r_{ij}})(\vec{m_j}\cdot\vec{r_{ij}})}{r_{ij}^5}\right)$: Nonuniform ion movement, such as piezoelectric distortions, alters dipolar fields, affecting magnetic anisotropy.
- Zeeman Energy $\left(\propto \vec{B}\cdot g_{\alpha\beta}\vec{S_i^\beta}\right)$: Electric fields modify the g-factor by altering electron wave functions or inducing local crystal field distortions, impacting magnetic moments.
- Frustrated Spin States: In type-II multiferroics (e.g., TbMnO₃), competing exchange interactions lead to frustrated spin configurations, such as cycloidal spirals. These non-collinear orders, driven by the DM interaction, break inversion symmetry, inducing electric polarization via $\vec{P} \propto \vec{e_{ij}} \times (\vec{S_i} \times \vec{S_j})$ [12, 13, 14]. Frustration enhances ME coupling by stabilizing complex magnetic structures sensitive to electric fields.

These interactions, modulated by electric fields, can differ across magnetized sublattices, enabling net ME effects in compensated antiferromagnets like $Cr_2O_3$.

### 2.3. Nanostructured Multiferroic Systems

Nanostructured multiferroics enable advanced magnetoelectric phenomena through intricate spin-lattice and spin-plasmon interactions. In layered metasurfaces, such as $SrTiO_3$/YIG multilayers, spin-orbit coupling and ferroelectric distortions create positive-negative birefringence. Electric and magnetic fields control refractive indices for different light polarizations, arising from the interplay of ferroelectric and ferromagnetic orders, which supports tunable photonic applications [15].

In nanostructured heterostructures, electric fields generate and amplify electromagnon solitons—hybrid magnon-phonon excitations. Spin-lattice coupling in the weakly nonlinear limit allows alternating electric fields to excite magnetic-like solitons near the magnetic band edge, achieving over 100-fold signal amplification. Nanoscale confinement enhances this dynamic ME effect, promising applications in high-frequency spintronic devices [16].

Plasmonic skyrmion lattices emerge in multiferroics like yttrium iron garnet on metallic substrates. ME coupling confines skyrmions—topologically protected spin textures—to plasmonic lattice sites formed by surface plasmon polariton interference. The plasmonic electric field interacts with magnetic spins, stabilizing these lattices and enabling optical control of collective magnonic modes for spintronics and sensing applications [17].

### 2.4. Quantum Approaches

Quantum mechanical approaches reveal intricate magnetoelectric interactions in multiferroic systems, leveraging quantum states and topological structures. In helical spin chains, such as $LiCu_2O_2$, electric field pulses enable high-fidelity transfer of single qubits and Bell pairs. Periodic "kicking" with electric fields mitigates impurities, ensuring robust quantum state transmission by modulating spin interactions, offering potential for quantum communication [18]. Quantum skyrmionic phases in two-dimensional helical spin lattices exhibit topological protection. Next-nearest-neighbor interactions enhance stability, while quenching from a skyrmionic to a ferromagnetic state induces a dynamical quantum phase transition (DQPT), marked by nonanalytic behavior in the rate function. This reflects entanglement dynamics, suggesting applications in quantum information processing [19]. Chiral multiferroic chains host prethermal Floquet time crystals (pFTCs), driven by periodic electric and magnetic fields. Disorder and Dzyaloshinskii-Moriya interactions stabilize chiral spin order, supporting long-lived coherent states. These systems function as quantum sensors for AC fields, with quantum Fisher information scaling superlinearly with spin number in the pFTC phase, surpassing the standard quantum limit. Tuning next-nearest-neighbor interactions and disorder optimizes sensor performance, leveraging many-body correlations and coherence [20].

## 3. Types of multiferroics

Multiferroic materials, defined by their simultaneous ferroelectric and magnetic ordering, are classified into single-phase and composite (multiphase) systems based on their structural and functional properties [5]. Single-phase multiferroics are further categorized into Type I, where ferroelectricity and magnetism arise independently, and Type II, where magnetism induces ferroelectricity [8]. Composite multiferroics, leveraging interfacial magnetoelectric (ME) coupling, encompass diverse architectures, including homogeneous mixtures, particulate composites, bi-layer and multi-layer heterostructures, and fiber-based structures [21, 22]. The following schematic (Fig. 1) illustrates this classification, highlighting the structural diversity.

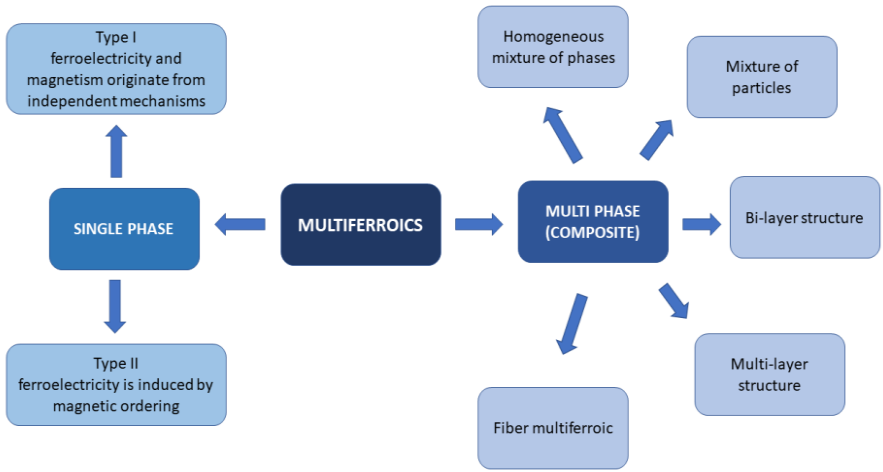

Fig. 1. Classification of multiferroic materials, illustrating single-phase and composite multiferroics.

### 3.1. Single-phase Multiferroics

Single-phase multiferroics are materials that exhibit at least two coexisting order parameters, typically ferromagnetism and ferroelectricity, within a single phase, resulting in uniform properties throughout the material. The coexistence of magnetic and electric ordering requires stringent crystal symmetry conditions, limiting the number of possible crystallographic structures. According to Shubnikov's classification, only 13 out of 122 crystallographic groups allow simultaneous magnetic and electric ordering, making single-phase multiferroics rare in nature and challenging to synthesize. These materials are broadly classified into two types: Type I multiferroics, such as $BiFeO_3$ (BFO), where ferroelectricity and magnetism originate from independent mechanisms, and Type II multiferroics, such as $TbMnO_3$, where ferroelectricity is induced by magnetic ordering, often resulting in stronger magnetoelectric (ME) coupling [7, 8, 10, 21]. For instance, $LiCuVO_4$, a Type II multiferroic, shows ferroelectric polarization induced by a helical spin order in the *ab* plane, with magnetic field-driven switching and significant polarization anisotropy (30 $\mu C/m^2$ at 1.95 K) [23]. These diverse mechanisms, spanning single-phase and composite systems, have driven significant advances in multiferroic research, enabling applications in sensors, memories, and energy technologies [8].

The ME coupling in these materials, which enables the control of magnetization by an electric field and, conversely, the manipulation of electric polarization by a magnetic field, is a key feature driving their potential applications. This coupling arises from various mechanisms, including domain wall interactions, exchange striction, and electromagnon excitations. For instance, domain walls in materials like $YMnO_3$ facilitate ME coupling through local symmetry breaking, as observed using nonlinear optics. Exchange striction in cycloidal multiferroics, such as orthorhombic rare-earth manganites ($RMn_2O_5$), induces ferroelectric polarization via lattice distortions driven by magnetic interactions. Electromagnons, resulting from the interaction between magnons and phonons, have been observed in materials like $TbMnO_3$ and $GdMnO_3$, contributing to ME coupling through spin-phonon interactions [24, 25].

While most single-phase multiferroics exhibit ME coupling at low temperatures, recent advances have identified materials with room-temperature properties, such as BFO, which shows a high ME coupling coefficient of approximately 3000 mV cm$^{-1}$ Oe$^{-1}$, and $Bi_5Ti_3FeO_{15}$, with a coefficient of 400 mV cm$^{-1}$ Oe$^{-1}$. The ability to synthesize $Bi_5Ti_3FeO_{15}$ without high-pressure techniques enhances its practical appeal compared to other perovskites. Recent studies have demonstrated enhanced polarization and magnetism in epitaxial BFO thin films, highlighting their potential for device applications. BFO exhibit enhanced polarization in thin film form due to epitaxial strain, with room-temperature spontaneous polarization reaching 50–60 $\mu C/cm^2$ [26].

### 3.2. Composite Multiferroics

Composite multiferroics are artificially engineered combinations of separate ferroelectric and magnetic materials, designed to achieve significantly higher magnetoelectric coupling coefficients compared to single-phase multiferroics, often at room temperature, which is critical for practical applications [21, 27, 28].

These materials can take the form of particulate composites, multilayer heterostructures, or self-assembled nanostructures, with the ME effect depending on the properties of individual phases, their microstructure, and the quality of the interfacial coupling [29]. Theoretical studies, demonstrate that in ferroelectric-ferromagnetic multilayers, changes in interfacial chemical bonding upon polarization reversal can induce significant magnetoelectric effects, enabling electric field control of magnetism [30]. The ability to control magnetization with an electric field and, conversely, manipulate electric polarization with a magnetic field makes composite multiferroics highly promising for applications in memory devices, sensors, and energy harvesting systems [10, 31].

The ME coupling in composite multiferroics arises from various mechanisms, including interfacial strain, charge-mediated coupling, and exchange bias. Interfacial strain coupling, observed in multilayer stacks combining magnetostrictive and piezoelectric phases, transfers strain induced by an external magnetic field to the piezoelectric phase, generating an electric charge. For example, layered composites like PZT-Terfenol-D exhibit ME coupling coefficients as high as 4.7 V cm$^{-1}$ Oe$^{-1}$, significantly surpassing particulate composites such as $CoFe_2O_4$-$BaTiO_3$ with a core-shell structure (8.1 mV cm$^{-1}$ Oe$^{-1}$) [29, 32]. Charge-mediated coupling, prominent in thin-film heterostructures like BFO-BTO, leverages spin-polarized charges at the ferroelectric-ferromagnetic interface to achieve ME coupling of approximately 61 mV cm$^{-1}$ Oe$^{-1}$ at room temperature, enabling voltage-controlled spintronic devices. Exchange bias, observed in systems like $Co_{0.9}Fe_{0.1}$-BFO, facilitates electric-field control of magnetization by exploiting uncompensated spins at the antiferromagnetic-ferromagnetic interface, with multilayer structures like Ta-Cu-$Mn_{70}$-$Ir_{30}$-$Fe_{70.2}Co_{7.8}Si_{12}B_{10}$ achieving ME coupling up to 96.7 V cm$^{-1}$ Oe$^{-1}$ at resonance frequencies [4]. Recent advances, such as terahertz-driven multiferroicity in $SrTiO_3$-based heterostructures, further expand the potential for tunable radio-frequency devices [33]. These developments highlight the potential of composite multiferroics for applications in magnetoelectric random-access memory, tunable radio-frequency devices, and energy harvesting technologies, as discussed in subsequent sections.

## 4. Applications of Multiferroics

Multiferroic materials, leveraging magnetoelectric coupling, enable a broad spectrum of applications by facilitating precise control of magnetic and electric properties, extending beyond the domains highlighted here. The following schematic (Fig. 2) illustrates key applications, including high-sensitivity magnetic field sensors, tunable radio-frequency devices, magnetoelectric random-access memory (MERAM), energy harvesters, quantum heat engines, and energy transport systems, which are briefly discussed in subsequent sections. These examples represent only a subset of the transformative potential of multiferroics in sensing, communication, data storage, and energy.

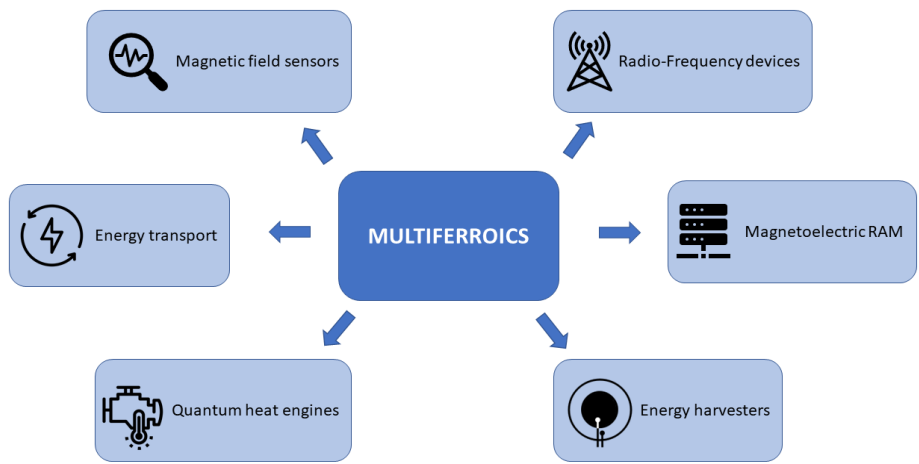

Fig. 2. Key applications of multiferroic materials

### 4.1. Magnetic Field Sensors

The magnetoelectric effect enables the development of highly sensitive magnetic field sensors with compact dimensions, leveraging the ability to couple magnetic and electric fields in multiferroic materials [5]. While both single-phase and composite multiferroics can theoretically be used, practical applications favor thin-film composite multiferroics due to their robust ME coupling at room temperature. These sensors typically utilize multilayer heterostructures, where ME coupling is mediated by mechanical strain or exchange bias, achieving high sensitivity and low detection limits [21, 27]. For instance, a sensor based on the Si-$SiO_2$-Pt-AlN-FeCoSiB heterostructure demonstrates a detection limit of 400 fT $Hz^{-1/2}$ at its resonance frequency, with picotesla-level sensitivity at nearby frequencies [34]. The structure of such sensors typically involves a multilayer composite, with a magnetostrictive layer coupled to a piezoelectric layer to maximize ME effects. Fig. 3a) illustrates a representative layer configuration for a magnetoelectric sensor, comprising a silicon substrate, magnetostrictive layer, lead seed layer, piezoelectric layer, and electrode, with the voltage signal measured between the electrode and magnetostrictive layer, enabling picotesla-level sensitivity.

Recent advancements focus on integrating multiferroic thin films into nanoelectromechanical systems (NEMS) resonators, which enable electrical detection of magnetic fields in the presence of a static magnetic field or electromechanical resonance frequency detection without a static field. Such innovations enhance the versatility of multiferroic sensors, making them suitable for applications in geomagnetic sensing, biomedical diagnostics, and navigation systems [31]. Ongoing research aims to further improve sensitivity by optimizing interfacial coupling in multilayer structures and exploring novel materials with enhanced ME coefficients at room temperature. These developments position multiferroic magnetic field sensors as critical components in next-generation sensing technologies [31, 34, 35].

### 4.2. Radio-Frequency Devices

The magnetoelectric effect enables precise control of high-frequency devices by switching ferroelectric polarization, offering a versatile platform for radio-frequency (RF) applications. Multiferroic materials, particularly composite thin-film heterostructures, facilitate the development of tunable RF components such as phase shifters, filters, oscillators, and miniature antennas, due to their ability to modulate magnetic properties with an electric field [5, 36]. These devices benefit from compact dimensions, enabling integration into electronic chips, and low power consumption required for polarization switching, which enhances energy efficiency. However, phase boundary noise in magnetostrictive layers can degrade performance, a challenge that can be mitigated by using materials like yttrium iron garnet (YIG) or barium titanate (BTO) [5, 33].

Recent advancements have focused on optimizing ME coupling in multilayer composites to achieve high tunability at resonance frequencies, making multiferroic RF devices suitable for applications in wireless communication and radar systems [36]. For example, heterostructures combining BTO with magnetostrictive alloys demonstrate enhanced frequency tunability, enabling compact antennas with improved performance. Ongoing research aims to further reduce noise and enhance ME coefficients at room temperature, leveraging materials like YIG-BTO composites to meet the demands of next-generation RF technologies [33]. These developments position multiferroic RF devices as critical components in advanced telecommunications and signal processing systems.

### 4.3. Magnetoelectric Random-Access Memory (MERAM)

Multiferroics enable advanced random-access memory (RAM) technologies by leveraging ME coupling to combine the advantages of ferroelectric RAM (FeRAM) and magnetic RAM (MRAM), offering low-power, non-volatile data storage [37]. Unlike conventional MRAM, which relies on high currents to generate local magnetic fields for switching ferromagnetic tunnel junctions, and FeRAM, which requires significant energy for polarization switching, ME-based RAM (MERAM) uses electric fields to control magnetization, significantly reducing energy consumption. This approach enables four-state logic in multiferroic tunnel junctions, with potential for up to eight states in complex structures, enhancing data storage density [5, 38]. The architecture of MERAM typically involves a multilayer heterostructure that combines ferroelectric-antiferromagnetic (FE-AFM) and ferromagnetic (FM) layers to achieve low-power, non-volatile data storage. Fig. 3b) illustrates a representative MERAM configuration, comprising an electrode, FE-AFM layer, two FM layers separated by a neutral layer, with binary information stored in the lower FM layer. Data writing is achieved by applying a voltage across the FE-AFM layer to switch magnetization, while readout relies on measuring resistance between the two FM layers.

Multiferroic heterostructures, such as Co-PZT-LSMO deposited on $SrTiO_3$, demonstrate efficient switching of magnetic states using short electric pulses at temperatures above 250 K, offering a pathway to low-power memory devices [38]. Recent advancements focus on antiferromagnetic MERAM (AF-MERAM), which eliminates ferromagnetic components to further enhance energy efficiency. For instance, thin films of $Cr_2O_3$ enable voltage-driven switching of antiferromagnetic order at room temperature, with write thresholds up to 50 times lower than traditional MERAM and fully electric readout, eliminating ferromagnetic hysteresis losses [39]. These developments, supported by optimized electrode materials and structural compatibility, position MERAM as a promising technology for next-generation non-volatile memory with applications in high-density data storage and neuromorphic computing [7, 37].

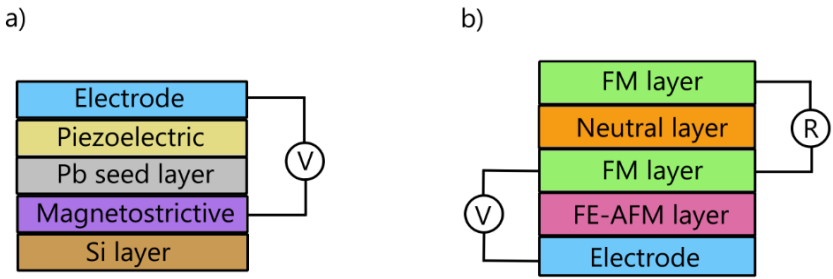

Fig. 3. **a)** Schematic of a multilayer magnetoelectric sensor for magnetic field detection, consisting of a silicon substrate, magnetostrictive layer, lead seed layer, piezoelectric layer, and electrode, with voltage measured between the electrode and magnetostrictive layer. **b)** Schematic of a magnetoelectric RAM (MERAM) heterostructure, consisting of an electrode, ferroelectric-antiferromagnetic (FE-AFM) layer, two ferromagnetic (FM) layers separated by a neutral layer, with voltage applied across FE-AFM for data writing and resistance measured between FM layers for readout.

### 4.4. Magnetoelectric Energy Harvesters

Magnetoelectric energy harvesters leverage the ME effect to convert magnetic field energy into electrical voltage, enabling compact, wireless devices for energy harvesting and sensing applications [27]. These devices are particularly promising for detecting weak magnetic fields and transmitting signals without external power sources, with significant potential in biomedical applications. For instance, a thin-film multiferroic heterostructure based on AlN-FeGaB, with dimensions of 250 × 175 μm², utilizes interfacial strain coupling to achieve high sensitivity [40]. The piezoelectric AlN layer converts polarization into mechanical strain, while the magnetostrictive FeGaB layer transforms magnetic field energy into mechanical strain, forming a nanoelectromechanical system (NEMS). This system operates at two resonance frequencies—2.51 GHz for receiving and transmitting electromagnetic signals and powering the device, and 63.6 MHz for detecting picotesla-level magnetic fields, such as the approximately 120 pT field generated by neuronal activity at a distance of 1 mm [40]. This enables contactless monitoring of neural activity, critical for implantable biomedical sensors operating at or above room temperature. Magnetoelectric energy harvesters typically employ a composite structure combining a piezoelectric layer with a magnetostrictive layer to convert magnetic field energy into electrical. Fig. 4a) illustrates this configuration, featuring a piezoelectric layer atop a magnetostrictive layer, with radio-frequency (RF) waves inducing magnetization changes in the magnetostrictive layer. This, via the magnetostrictive effect, deforms the piezoelectric layer, generating a voltage for data readout. Conversely, applying a voltage deforms the piezoelectric layer, triggering RF wave emission for data transmission.

Multiferroic photovoltaics represent another avenue for energy harvesting, utilizing materials like BFO with low bandgap energies to achieve efficient charge separation and high power conversion efficiency compared to traditional ferroelectrics. The ME effect enhances the separation of electron-hole pairs, making multiferroic photovoltaics promising for solar energy applications [4, 5]. Ongoing research focuses on optimizing ME coupling at room temperature and developing biocompatible materials to expand the use of multiferroic harvesters in wearable devices, wireless sensor networks, and sustainable energy systems [41]. These advancements highlight the versatility of multiferroic energy harvesters in addressing both biomedical and renewable energy challenges.

### 4.5. Quantum Heat Engines

Multiferroic materials enable quantum heat engines, such as the quantum Otto cycle, to convert electrical energy into mechanical work via magnetoelectric (ME) coupling. In chiral multiferroic chains, like $LiCu_2O_2$, electric fields modulate spin interactions and quantum energy states, enhancing thermodynamic cycle performance. Fig. 4b) illustrates the thermodynamic cycle of such an engine, depicting a chiral multiferroic chain interacting with a cold bath and a hot bath through alternating adiabatic and isochoric processes, enabling efficient energy conversion driven by spin entanglement and electric field control. Unlike classical engines limited by Carnot efficiency, these systems exploit spin entanglement and coherence, achieving high efficiencies under idealized conditions. Three-spin chains demonstrate near 100% efficiency due to minimized entropic losses and helical spin asymmetry, amplifying quantum interference effects [39, 42].

Finite-time quantum thermodynamic cycles in chiral multiferroics utilize shortcuts to adiabaticity to minimize adiabatic stroke duration, ensuring high output power. Electric field pulses control noncollinear spin order, enabling reversible cycles with zero irreversible work at stroke completion and efficiencies up to 47%. Thermal relaxation, modeled via Lindblad master equations, reduces efficiency to 23% under Gibbs ensemble thermalization, highlighting coherence's role in performance optimization [43].

Plasmonic skyrmion lattices in multiferroics leverage topologically protected skyrmions confined by plasmonic modes. These engines eliminate quantum friction—irreversible work from inter-level transitions—due to near-zero transition matrix elements, achieving efficiencies above 45% without adiabatic shortcuts. Electric fields manipulate skyrmion numbers, precisely controlling output power. The Dzyaloshinskii-Moriya interaction, intrinsic to ME coupling, stabilizes skyrmions, making them robust at finite temperatures in materials like yttrium iron garnet [19].

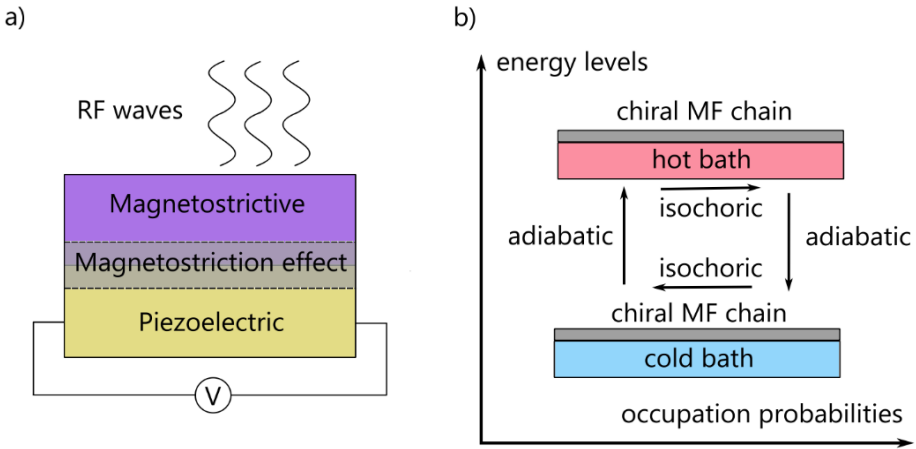

Fig. 4. **a)** Schematic of a magnetoelectric energy harvester, comprising a piezoelectric layer on a magnetostrictive layer, with RF waves inducing magnetization changes for voltage generation (data readout) and voltage-induced deformation triggering RF emission (data transmission). **b)** Schematic of a quantum Otto cycle in a chiral multiferroic chain, illustrating the thermodynamic cycle with adiabatic and isochoric processes connecting a cold bath and a hot bath for efficient energy conversion.

### 4.6. Energy Transport in Multiferroics

Multiferroic composites enable efficient energy transport by leveraging ME coupling to control signal propagation and heat transfer across ferroelectric-ferromagnetic interfaces. In systems like $BaTiO_3$ combined with ferromagnetic materials, theoretical studies reveal a signal amplitude threshold above which the interface becomes transparent to ferroelectric modes, allowing excitations to penetrate the ferromagnetic layer [44]. This transparency, governed by the strength of ME coupling, facilitates precise signal transmission, with potential applications in optoelectronic devices and next-generation memory technologies [5]. Intriguingly, sub-threshold signals can be amplified by controlled noise levels, resembling stochastic resonance, which enhances system sensitivity and signal amplification under real-world conditions.

ME coupling also enables the development of thermal diodes, which provide directional heat transport analogous to electrical diodes, as demonstrated in composites of tetragonal iron and $BaTiO_3$. These diodes exploit phonon-mediated heat transfer across a broad terahertz spectrum, with asymmetric heat flow driven by temperature gradients and modulated by ferroelectric polarization and ferromagnetic domain orientation [45]. External electric fields enhance thermal conductivity and rectification, while magnetic fields suppress heat flux, offering dynamic control over thermal transport. These properties position multiferroic thermal diodes as key components in thermal management technologies, such as heat switches, thermal memories, and cooling systems for nanoelectronics. Ongoing research aims to optimize interface dynamics and material properties to enhance energy transport efficiency, paving the way for advanced sensors, integrated circuits, and energy-efficient nanodevices [8].

## 5. Conclusion

Multiferroic materials, defined by their ability to couple magnetic and electric fields, provide a versatile platform for advanced technologies. Single-phase multiferroics, like BFO, exhibit intrinsic ME coupling but face limitations in room-temperature performance, whereas composite multiferroics, leveraging interfacial strain, charge, or exchange bias, achieve higher ME coefficients for practical applications. This review explores ME phenomena through phenomenological, microscopic, nanostructured, and quantum approaches, revealing mechanisms from macroscopic free energy models to spin-lattice interactions, plasmonic skyrmion lattices, and quantum entanglement-driven energy conversion.

Applications span high-sensitivity magnetic field sensors, tunable radio-frequency devices, energy-efficient magnetoelectric random-access memory (MERAM), energy harvesters, quantum heat engines, and thermal diodes. Composite multiferroic sensors, such as $Si$-$SiO_2$-$Pt$-$AlN$-$FeCoSiB$, achieve picotesla-level sensitivity for biomedical diagnostics and navigation. Radio-frequency devices utilize ME coupling for low-power antennas and filters, while MERAM enables non-volatile, high-density data storage. Energy harvesters, like AlN-FeGaB neural sensors, convert ambient energy into electrical signals, supporting sustainable technologies. Nanostructured systems, including plasmonic skyrmion lattices and metasurfaces, enhance ME effects for optical and spintronic applications, while quantum heat engines in chiral multiferroics, like $LiCu_2O_2$, leverage spin entanglement and topological protection for efficient nanoscale energy conversion.

Future research should focus on optimizing room-temperature ME coupling, developing biocompatible and scalable materials, and addressing interface noise, quantum decoherence, and coherence stability in nanostructured and quantum systems. Advances in thin-film engineering and nanotechnology will drive the integration of multiferroics into devices for neuromorphic computing, quantum technologies, and sustainable energy, bridging fundamental science and engineering to meet global technological demands.

# Literature

[1] I. E. Dzyaloshinski, "On the Magneto-Electrical Effect in Antiferromagnets," *Soviet Physics JETP,* vol. 10, p. 628–629, 1959.

[2] D. N. Astrov, "The Magnetoelectric Effect in Antiferromagnets," *Soviet Physics JETP,* vol. 11, p. 708–709, 1960.

[3] V. J. Folen, G. T. Rado and E. W. Stalder, "Anisotropy of the Magnetoelectric Effect in Cr_2O_3," *Physical Review Letters,* vol. 6, p. 607–608, 1961.

[4] R. Gupta and R. K. Kotnala, "A review on current status and mechanisms of room-temperature magnetoelectric coupling in multiferroics for device applications," *Journal of Materials Science,* vol. 57, p. 12710–12737, 2022.

[5] R. Ramesh and N. A. Spaldin, "Multiferroics: progress and prospects in thin films," *Nature materials,* vol. 6, p. 21–29, 2007.

[6] N. A. Spaldin and M. Fiebig, "The renaissance of magnetoelectric multiferroics," *Science,* vol. 309, p. 391–392, 2005.

[7] K. F. Wang, J.-M. Liu and Z. F. Ren, "Multiferroicity: the coupling between magnetic and polarization orders," *Advances in Physics,* vol. 58, p. 321–448, 2009.

[8] M. Fiebig, T. Lottermoser, D. Meier and M. Trassin, "The evolution of multiferroics," *Nature Reviews Materials,* vol. 1, p. 1–14, 2016.

[9] D. I. Khomskii, "Multiferroics: Different ways to combine magnetism and ferroelectricity," *Journal of Magnetism and Magnetic Materials,* vol. 306, p. 1–8, 2006.

[10] M. Fiebig, "Revival of the magnetoelectric effect," *Journal of physics D: applied physics,* vol. 38, p. R123, 2005.

[11] S.-W. Cheong and M. Mostovoy, "Multiferroics: a magnetic twist for ferroelectricity," *Nature materials,* vol. 6, p. 13–20, 2007.

[12] S. Park, Y. J. Choi, C. L. Zhang and S.-W. Cheong, "Ferroelectricity in an S= 1/2 chain cuprate," *Physical review letters,* vol. 98, p. 057601, 2007.

[13] M. Mostovoy, "Ferroelectricity in spiral magnets," *Physical review letters,* vol. 96, no. 6, p. 067601, 2006.

[14] H. Katsura, N. Nagaosa and A. V. Balatsky, "Spin current and magnetoelectric effect in noncollinear magnets," *Physical review letters,* vol. 95, p. 057205, 2005.

[15] R. Khomeriki, L. Chotorlishvili, I. Tralle and J. Berakdar, "Positive--negative birefringence in multiferroic layered metasurfaces," *Nano Letters,* vol. 16, no. 11, pp. 7290--7294, 2016.

[16] R. Khomeriki, L. Chotorlishvili, B. A. Malomed and J. Berakdar, "Creation and amplification of electromagnon solitons by electric field in nanostructured multiferroics," *Physical Review B,* vol. 91, p. 041408, 2015.

[17] X.-G. Wang, L. Chotorlishvili, N. Arnold, V. K. Dugaev, I. Maznichenko, J. Barnaś, P. A. Buczek, S. S. P. Parkin and A. Ernst, "Plasmonic skyrmion lattice based on the magnetoelectric effect," *Physical Review Letters,* vol. 125, p. 227201, 2020.

[18] H. Verma, L. Chotorlishvili, J. Berakdar and S. K. Mishra, "Qubit (s) transfer in helical spin chains," *Europhysics Letters,* vol. 119, p. 30001, 2017.

Magnetoelectric Multiferroics: From Fundamentals to Transformative Applications    17[19] V. Vijayan, L. Chotorlishvili, A. Ernst, M. I. Katsnelson, S. S. P. Parkin and S. K. Mishra, "Quantum heat engine with near-zero irreversible work utilizing quantum skyrmion working substance," *Physica A: Statistical Mechanics and its Applications,* p. 130599, 2025.

[20] R. K. Shukla, L. Chotorlishvili, S. K. Mishra and F. Iemini, "Prethermal Floquet time crystals in chiral multiferroic chains and applications as quantum sensors of AC fields," *Physical Review B,* vol. 111, p. 024315, 2025.

[21] W. Eerenstein, N. D. Mathur and J. F. Scott, "Multiferroic and magnetoelectric materials," *nature,* vol. 442, p. 759–765, 2006.

[22] M. M. Vopson, "Fundamentals of multiferroic materials and their possible applications," *Critical Reviews in Solid State and Materials Sciences,* vol. 40, p. 223–250, 2015.

[23] F. Schrettle, S. Krohns, P. Lunkenheimer, J. Hemberger, N. Büttgen, H.-A. Krug von Nidda, A. V. Prokofiev and A. Loidl, "Switching the ferroelectric polarization in the S= 1/ 2 chain cuprate Li Cu VO 4 by external magnetic fields," *Physical Review B—Condensed Matter and Materials Physics,* vol. 77, p. 144101, 2008.

[24] M. Fiebig, T. Lottermoser, D. Fröhlich, A. V. Goltsev and R. V. Pisarev, "Observation of coupled magnetic and electric domains," *Nature,* vol. 419, p. 818–820, 2002.

[25] D. Meier, M. Maringer, T. Lottermoser, P. Becker, L. Bohatỳ and M. Fiebig, "Observation and coupling of domains in a spin-spiral multiferroic," *Physical review letters,* vol. 102, p. 107202, 2009.

[26] J. B. N. J. Wang, J. B. Neaton, H. Zheng, V. Nagarajan, S. B. Ogale, B. Liu, D. Viehland, V. Vaithyanathan, D. G. Schlom, U. V. Waghmare and others, "Epitaxial BiFeO3 multiferroic thin film heterostructures," *science,* vol. 299, p. 1719–1722, 2003.

[27] C.-W. Nan, M. I. Bichurin, S. Dong, D. Viehland and G. Srinivasan, "Multiferroic magnetoelectric composites: Historical perspective, status, and future directions," *Journal of applied physics,* vol. 103, 2008.

[28] L. Chotorlishvili, C. Jia, D. A. Rata, L. Brandt, G. Woltersdorf and J. Berakdar, "Magnonic Magnetoelectric Coupling in

18 M. Wanic

Ferroelectric/Ferromagnetic Composites," *physica status solidi (b),* vol. 257, p. 1900750, 2020.

[29] Y. Wang, J. Hu, Y. Lin and C.-W. Nan, "Multiferroic magnetoelectric composite nanostructures," *NPG asia materials,* vol. 2, p. 61–68, 2010.

[30] C.-G. Duan, S. S. Jaswal and E. Y. Tsymbal, "Predicted Magnetoelectric Effect in Fe/BaTiO 3 Multilayers:<? format?> Ferroelectric Control of Magnetism," *Physical Review Letters,* vol. 97, p. 047201, 2006.

[31] N. A. Spaldin and R. Ramesh, "Advances in magnetoelectric multiferroics," *Nature materials,* vol. 18, p. 203–212, 2019.

[32] A. Chaudhuri and K. Mandal, "Large magnetoelectric properties in CoFe2O4: BaTiO3 core–shell nanocomposites," *Journal of Magnetism and Magnetic Materials,* vol. 377, p. 441–445, 2015.

[33] M. Basini, M. Pancaldi, B. Wehinger, M. Udina, V. Unikandanunni, T. Tadano, M. C. Hoffmann, A. V. Balatsky and S. Bonetti, "Terahertz electric-field-driven dynamical multiferroicity in SrTiO3," *Nature,* vol. 628, p. 534–539, 2024.

[34] E. Yarar, S. Salzer, V. Hrkac, A. Piorra, M. Höft, R. Knöchel, L. Kienle and E. Quandt, "Inverse bilayer magnetoelectric thin film sensor," *Applied Physics Letters,* vol. 109, 2016.

[35] Y.-H. Chu, L. W. Martin, M. B. Holcomb, M. Gajek, S.-J. Han, Q. He, N. Balke, C.-H. Yang, D. Lee, W. Hu and others, "Electric-field control of local ferromagnetism using a magnetoelectric multiferroic," *Nature materials,* vol. 7, p. 478–482, 2008.

[36] Y. K. Fetisov, V. L. Preobrazhenskii and P. Pernod, "Bistability in a nonlinear magnetoacoustic resonator," *Journal of communications technology and electronics,* vol. 51, p. 218–230, 2006.

[37] M. Bibes and A. Barthélémy, "Towards a magnetoelectric memory," *Nature materials,* vol. 7, p. 425–426, 2008.

[38] D. Pantel, S. Goetze, D. Hesse and M. Alexe, "Reversible electrical switching of spin polarization in multiferroic tunnel junctions," *Nature materials,* vol. 11, p. 289–293, 2012.

[39] T. Kosub, M. Kopte, R. Hühne, P. Appel, B. Shields, P. Maletinsky, R. Hübner, M. O. Liedke, J. Fassbender, O. G. Schmidt and others, "Purely antiferromagnetic magnetoelectric random access memory," *Nature communications,* vol. 8, p. 13985, 2017.

[40] M. Zaeimbashi, M. Nasrollahpour, A. Khalifa, A. Romano, X. Liang, H. Chen, N. Sun, A. Matyushov, H. Lin, C. Dong and others, "Ultra-compact dual-band smart NEMS magnetoelectric antennas for simultaneous wireless energy harvesting and magnetic field sensing," *Nature communications,* vol. 12, p. 3141, 2021.

[41] V. Annapureddy, H. Palneedi, G.-T. Hwang, M. Peddigari, D.-Y. Jeong, W.-H. Yoon, K.-H. Kim and J. Ryu, "Magnetic energy harvesting with magnetoelectrics: an emerging technology for self-powered autonomous systems," *Sustainable Energy & Fuels,* vol. 1, p. 2039–2052, 2017.

[42] M. Azimi, L. Chotorlishvili, S. K. Mishra, S. Greschner, T. Vekua and J. Berakdar, "Helical multiferroics for electric field controlled quantum information processing," *Physical Review B,* vol. 89, p. 024424, 2014.

[43] L. Chotorlishvili, M. Azimi, S. Stagraczyński, Z. Toklikishvili, M. Schüler and J. Berakdar, "Superadiabatic quantum heat engine with a multiferroic working medium," *Physical Review E,* vol. 94, p. 032116, 2016.

[44] L. Chotorlishvili, R. Khomeriki, A. Sukhov, S. Ruffo and J. Berakdar, "Dynamics of localized modes in a composite multiferroic chain," *Physical Review Letters,* vol. 111, p. 117202, 2013.

[45] L. Chotorlishvili, S. R. Etesami, J. Berakdar, R. Khomeriki and J. Ren, "Electromagnetically controlled multiferroic thermal diode," *Physical Review B,* vol. 92, p. 134424, 2015.